\documentclass[aps,pra,twocolumn,superscriptaddress,nofootinbib]{revtex4-2}
\usepackage{amsmath}
\usepackage{amssymb}
\usepackage{graphicx}

\begin{document}

\title{Quantum evolution with classical fields}

\author{Christof Wetterich}
\affiliation{Institut f\"ur Theoretische Physik, Universit\"at Heidelberg, Philosophenweg 16, D-69120 Heidelberg}

\begin{abstract}
Wave guides for classical electromagnetic fields can realize the discrete quantum evolution of the wave function for a system of qubits. Phase shifts, switches and beam splits allow for the construction of arbitrary quantum gates. They can act at once on a large number of qubits. For this correlation based photonic quantum computer the channels of the wave guides represent basis states of a multi-qubit system rather than individual qubits. 
With probabilistic initial conditions for a large number of channels the classical fields in wave guides realize the density matrix for a probabilistic quantum computer obeying the von Neumann equation.
\end{abstract}

\maketitle

\nocite{LASO,BRSO,SHE,WETZ,Giordani2023,Shastri2021,Jorgensen2022,Bruckerhoff2024,Jung2025,Sarwat2025,Reck1994,DiVincenzo1998,DiVincenzo2000,OBrien2007,Politi2008,Matthews2009,Zheng2013,Romero2024,PerezGarcia2015,deOliveira2005,Marques2012,Zhang2010,Zhang2012,MachoOrtiz2023,Spreeuw1998,Dragoman2004,Longhi2011,CWQMCS,CWEQM,Wetterich2018a,Wetterich2018b,ProbabilisticWorld,ProbabilisticWorldII,Wetterich2010,CWL,CWQO,CWDS,CWTE,Pehle2018,Pehle2022,Wetterich2025,Sexty2018,Bell1964,Clauser1969,Wetterich2021,Wetterich2022,Kreuzkamp2024}

\section{Introduction}

The propagation of classical electromagnetic fields in wave guides with phase shifts, switches or beam splits may lead to new forms of computing \cite{LASO,BRSO,SHE,WETZ}. For optical circuits with a large number of channels the preparation of a precise initial state may become difficult. It gets replaced by a probability distribution over initial classical field configurations. This will lead to photonic forms of probabilistic computing. We investigate here the question if optical circuits can perform the operations of quantum gates in a quantum computer. We find that a large class of systems in linear optics can be described by the Schr\"odinger equation for a complex wave function, or its discrete version by a sequence of unitary quantum operations. In case of probabilistic initial conditions this gets replaced by the von-Neumann equation for the density matrix.

There has been rapid recent progress in the field of optical circuits \cite{Giordani2023,Shastri2021,Jorgensen2022,Bruckerhoff2024,Jung2025,Sarwat2025}. One can profit from many developments which use quantum photons as qubits (``photonic qubits'') \cite{Reck1994,DiVincenzo1998,DiVincenzo2000,OBrien2007,Politi2008,Matthews2009,Zheng2013,Romero2024}, employ classical electromagnetic fields for the realization of computational tasks as the Deutsch-Jozsa algorithm \cite{PerezGarcia2015,deOliveira2005,Marques2012,Zhang2010,Zhang2012} or propose generalized forms of computing \cite{MachoOrtiz2023}.
While one can recognize certain quantum structures in these developments \cite{Spreeuw1998,Dragoman2004,Longhi2011}, the present proposal goes further. We argue that the evolution of a quantum system with all its properties can be realized by classical electromagnetic fields. We do not use quantum photons.

On the conceptual side it has been claimed that certain quantum systems can be obtained from classical probabilistic systems \cite{CWQMCS,CWEQM,Wetterich2018a,Wetterich2018b,ProbabilisticWorld,ProbabilisticWorldII}. An embedding of quantum mechanics in classical statistics washes out sharp boundaries between quantum and classical systems. One may therefore aim to find in practice classical probabilistic systems which follow a quantum evolution. We propose here the realization of a type of quantum computer by classical electromagnetic fields in wave guides. It emulates a quantum system by representing probability amplitudes by classical electromagnetic fields, and probabilities by intensities which are quadratic in the fields.

Classical probabilistic systems are characterized by a probability distribution. This may specify probabilities for events at all times. A key question is the time-evolution of the probabilistic information of a time-local subsystem. Such a subsystem ``integrates out'' the detailed probabilistic information about past and future events. It only retains the probabilistic information relevant for a given time $t$. Instead of using only time-dependent probabilities $p_\tau(t)$, a better encoding of the probabilistic information of the subsystem employs classical wave functions \cite{Wetterich2010,CWQMCS,Wetterich2018a,Wetterich2018b} with real components $q_\tau(t)$, or a generalization to classical density matrices $\rho_{\tau\rho}(t)$. In the context of this note the components of the classical wave functions are given by the square root of the probabilities up to a sign, $p_\tau(t) = q_\tau^2(t)$. The normalization of the probabilities implies that these wave functions are unit vectors, $\sum_\tau q_\tau^2(t) = 1$. Time evolution is a rotation of the unit vectors. If the evolution is linear, no information is lost. A complex structure is a map from the real wave function to a complex wave function. If the complex structure is compatible with the evolution, the rotation of the real vector $q$ is mapped to a unitary evolution of the complex wave function -- this is precisely the structure of the evolution in quantum mechanics. This structure is rather general, and it becomes a question in which fields one can find practical realizations for the linear time evolution of the probability amplitudes $q_\tau(t)$. 

Recent examples are classical field theories with probabilistic initial conditions \cite{CWL}. In the presence of a conserved charge they contain a single-particle subsystem whose time evolution follows the Schr\"odinger equation for a quantum particle in a potential \cite{CWDS}. The quantum formalism for classical statistics \cite{Wetterich2018b} extends to a wider class of transport equations with probabilistic initial conditions \cite{CWTE} and even to probabilistic classical particles \cite{CWQO}. We ask here if one can find simple systems which emulate directly the probability amplitudes whose square yields the probabilities. Classical electromagnetic fields in wave guides are interesting candidates for this purpose. The probabilities can be identified with relative intensities, which are quadratic in the electromagnetic fields. This suggests to use electromagnetic fields as probability amplitudes or classical wave functions. Within linear optics the superposition principle holds and the evolution of the wave function is linear.

Besides possible practical applications for computing we are interested in the general structure how classical fields can represent probability amplitudes with a linear evolution. After introducing a simple complex structure we discuss a correlation based photonic quantum computer which can perform arbitrary unitary transformations on a system of qubits. We only use classical electromagnetic fields in wave guides, in contrast to interesting ideas of using quantum photons as qubits. The wave guides will be associated to basis states of the qubit system, rather than to qubits themselves. We keep the discussion very basic, with the aim to demonstrate that nothing is missing in the realization of a quantum system by classical fields.

From the point of view of the microscopic system of electromagnetic fields the quantum system is a probabilistic subsystem. For the subsystem only a small part of the microscopic information is available. This information is sufficient, however, to determine expectation values of ``statistical observables'' which do not obey Bell's inequalities. Other examples for such probabilistic subsystems have been found in artificial neural networks \cite{Pehle2018} or neuromorphic computing \cite{Pehle2022}.

\section{Probabilities and classical wave function}

Consider $N$ classical real field variables $F_\tau$. Such variables are typically related to suitable mean values of components of electromagnetic fields in a wave guide. The partial intensities in these field variables are $I_\tau = F_\tau^2$. Relative intensities are associated to probabilities $p_\tau$. They obtain by normalizing, with total intensity $I_{\rm tot}$,
\begin{equation}
I_{\rm tot} = \sum_\tau I_\tau\,, \quad p_\tau = \frac{I_\tau}{I_{\rm tot}}\,. \label{eq:1}
\end{equation}
The ensemble of relative intensities has the properties of a probability distribution,
\begin{equation}
p_\tau \ge 0\,, \quad \sum_\tau p_\tau = 1\,. \label{eq:2}
\end{equation}
The components of a classical wave function $q$ are the normalized field variables
\begin{equation}
q_\tau = \frac{F_\tau}{\sqrt{I_{\rm tot}}}\,, \quad q_\tau^2 = p_\tau\,. \label{eq:3}
\end{equation}
They are fixed by the probabilities up to a sign.

The evolution of the fields after a finite time step $\varepsilon$ can be written in the form
\begin{equation}
q_\tau(t+\varepsilon) = A_{\tau\rho}(t;q) q_\rho(t)\,. \label{eq:4}
\end{equation}
(We sum over repeated indices, $A_{\tau\rho} q_\rho \equiv \sum_\rho A_{\tau\rho} q_\rho$.) In the absence of dissipation the total intensity is conserved, $I_{\rm tot}(t+\varepsilon) = I_{\rm tot}(t)$. This implies that the normalization of the probabilities is preserved
\begin{equation}
\sum_\tau p_\tau(t+\varepsilon) = \sum_\tau p_\tau(t) = 1\,, \label{eq:5}
\end{equation}
and $q$ is therefore a unit vector at all times, $\sum_\tau q_\tau^2(t+\varepsilon) = \sum_\tau q_\tau^2(t)$. As a direct consequence, $A$ is an orthogonal matrix
\begin{equation}
A^T(t;q) A(t;q) = 1\,. \label{eq:6}
\end{equation}

If the evolution of the field components obeys a linear evolution law, as given, for example, by the linear Maxwell equations in medium, the matrix $A$ does not depend on $q$. It can be identified with the step evolution operator $\hat S$ of a linear evolution
\begin{equation}
q(t+\varepsilon) = \hat S(t) q(t)\,. \label{eq:7}
\end{equation}
An arbitrary orthogonal matrix $\hat S$ can be written in terms of a hermitian Hamiltonian $\bar H(t)$,
\begin{equation}
\hat S(t) = \exp(-i\varepsilon \bar H(t))\,, \quad \bar H^\dagger = \bar H\,. \label{eq:8}
\end{equation}
For a linear evolution $\bar H(t)$ is field-independent, but it may depend on the time step labeled by $t$. Taking $\bar H(t)$ to be constant between $t$ and $t+\varepsilon$ we can interpret eq.~(\ref{eq:8}) as a continuous evolution in time with piecewise constant Hamiltonian. On the other hand, if the Hamiltonian $H(t)$ for a continuous time evolution is known, as typically for Maxwell's equations in matter, one has $\bar H(t) = \frac{1}{\varepsilon}\int_t^{t+\varepsilon} dt'\, H(t')$. At this stage our system describes the evolution of a generalized quantum system in a real and discrete setting. In particular, the linearity of the evolution equation (\ref{eq:7}) entails the superposition principle for possible solutions.

\section{Complex structure and unitary evolution}

A complex structure maps the real wave function $q$ to a complex wave function $\psi$.
As a first example we take an evolution for which the field variables $F_\tau$ can be organized in pairs for which the two members rotate into each other. This means that the sum of the two partial intensities does not depend on time,
\begin{equation}
\tau = (\alpha,\eta)=\alpha_\eta\,, \quad \eta = (1,2)\,,
\end{equation}
\begin{equation}
I_{\alpha_1}(t+\varepsilon) + I_{\alpha_2}(t+\varepsilon) = I_{\alpha_1}(t) + I_{\alpha_2}(t)\,. \label{eq:9}
\end{equation}
With $q_{\alpha_1}^2(t+\varepsilon) + q_{\alpha_2}^2(t+\varepsilon) = q_{\alpha_1}^2(t) + q_{\alpha_2}^2(t)$ one has a block diagonal form of the evolution operator $\hat S(t)$, where each block is a rotation in a two-dimensional subspace
\begin{align}
q_{\alpha_1}(t+\varepsilon) &= \cos\gamma_\alpha(t) q_{\alpha_1}(t) + \sin\gamma_\alpha(t) q_{\alpha_2}(t)\,,\notag\\
q_{\alpha_2}(t+\varepsilon) &= -\sin\gamma_\alpha(t) q_{\alpha_1}(t) + \cos\gamma_\alpha(t) q_{\alpha_2}(t)\,. \label{eq:10}
\end{align}
We can combine the pair of real components $q_{\alpha_1}$ and $q_{\alpha_2}$ into the component of a complex wave function $\psi$,
\begin{equation}
\psi_\alpha = q_{\alpha_1} + i q_{\alpha_2}\,, \quad \psi^\dagger\psi = \sum_\alpha \psi_\alpha^* \psi_\alpha = 1\,, \label{eq:11}
\end{equation}
where the number of components is $N/2$, $\alpha = 1,\dots,N/2$. The evolution is given by a unitary operator
\begin{equation}
\psi_\alpha(t+\varepsilon) = U_{\alpha\beta}(t)\psi_\beta(t)\,, \quad U^\dagger(t) U(t) = 1\,. \label{eq:12}
\end{equation}
For our simple example $U$ is a diagonal matrix, $U = {\rm diag}(e^{-i\gamma_\alpha})$, with diagonal Hamiltonian $\bar H = {\rm diag}(\gamma_\alpha/\varepsilon)$.

More generally, a complex structure is a map $q \to \psi$ which is defined by the presence of a pair $(K,I)$ of anticommuting discrete transformations acting on $q$, $K^2=1$, $I^2=-1$, $\{K,I\}=0$. In the complex picture $K$ becomes complex conjugation and $I$ multiplication by $i$. If the evolution is compatible with a given complex structure, an orthogonal evolution operator in the real picture is represented by a unitary evolution operator in the complex picture. Thus eq.~(\ref{eq:12}) holds for all possible complex structures which are compatible with the evolution \cite{ProbabilisticWorld,Wetterich2025}.

\section{Qubits}

Let us take $N = 2^{M_q+1}$. The $2^{M_q}$ components of the complex wave function $\psi$ can be associated with the $2^{M_q}$ basis states for $M_q$ qubits. We may define this basis by eigenstates of the $z$-components $S_j^{(z)}$ of the spins of the qubits, $s_j = 2S_j^{(z)}/\hbar$, $j=1,\dots,M_q$, $s_j = \pm 1$. For the example $N=16$, $M_q=3$ we associate $\alpha = 1,\dots,8$ to the states $(1,1,1)$, $(1,1,0)$, $(1,0,1)$, $(1,0,0)$, $(0,1,1)$, $(0,1,0)$, $(0,0,1)$, $(0,0,0)$, where $(1,0,1)$ stands for $s_1=1$, $s_2=-1$, $s_3=1$. In this notation the analogy to bits or fermionic occupation numbers $n_j = (s_j+1)/2$ is apparent. In this setting $\bar p_\alpha = |\psi_\alpha|^2= q^2_{\alpha_1}+q^2_{\alpha_2} = p_{\alpha_1}+p_{\alpha_1}$ denotes the probability to find the three qubit system in the state $\alpha$, e.g.\ $|\psi_3|^2$ is the probability to find $s_1=1$, $s_2=-1$, $s_3=1$. 

The expectation values of $s_j$ are directly related to the probabilities $\bar p_\alpha$,
\begin{align}
\langle s_1 \rangle &= \bar p_1+\bar p_2+\bar p_3+\bar p_4 - (\bar p_5+\bar p_6+\bar p_7+\bar p_8)\,,\notag\\
\langle s_2 \rangle &= \bar p_1+\bar p_2+\bar p_5+\bar p_6 - (\bar p_3+\bar p_4+\bar p_7+\bar p_8)\,,\notag\\
\langle s_3 \rangle &= \bar p_1+\bar p_3+\bar p_5+\bar p_7 - (\bar p_2+\bar p_4+\bar p_6+\bar p_8)\,. \label{eq:13}
\end{align}
Correlations are represented by the $\bar p_\alpha$ as well, as
\begin{align}
\langle s_1 s_2\rangle &= \bar p_1+\bar p_2+\bar p_7+\bar p_8-(\bar p_3+\bar p_4+\bar p_5+\bar p_6)\,,\notag\\
\langle s_1 s_3\rangle &= \bar p_1+\bar p_3+\bar p_6+\bar p_8-(\bar p_2+\bar p_4+\bar p_5+\bar p_7)\,. \label{eq:14}
\end{align}
We recall here that $\bar p$ is a sum of two partial relative intensities $\bar p_\alpha = p_{\alpha,1}+p_{\alpha,2}$. These expectation values can be measured in a simple way by comparing relative intensities in the sums of various wave guides.

As a simple realization we consider $2^{M_q}$ independent identical waveguides. Taking the $z$-coordinate in the direction of the waveguide and assuming for each waveguide two field components rotating into each other, one has for a given waveguide $\alpha$ the three-dimensional electromagnetic field components $\vec{\mathcal F}_{1,2}^{(\alpha)}(x,y,z,t)$,
\begin{align}
\vec{\mathcal F}_1 &= \vec f_1(x,y)\cos(\beta z - \omega t) - \vec f_2(x,y)\sin(\beta z - \omega t)\,,\notag\\
\vec{\mathcal F}_2 &= \vec f_2(x,y)\cos(\beta z - \omega t) + \vec f_1(x,y)\sin(\beta z - \omega t)\,. \label{eq:15}
\end{align}
Here $\vec{\mathcal F}_1^2 + \vec{\mathcal F}_2^2 = \vec f_1^{\,2}(x,y) + \vec f_2^{\,2}(x,y)$ is proportional to the partial intensity $I_\alpha$ of the given waveguide, that we may define here by
\begin{equation}
I_\alpha = \int_{x,y} \left(\vec f_1^{\,2} + \vec f_2^{\,2}\right)\,. \label{eq:15A}
\end{equation}

The component $\psi_\alpha(t)$ of the complex wave function is determined by the relative intensity $p_\alpha(t)$ and a phase $\varphi_\alpha(t)$, $\psi_\alpha(t) = p_\alpha(t) e^{i\varphi_\alpha(t)}$. For the phase we note the identities $\vec{\mathcal F}_1 = {\rm Re}\left[(\vec f_1+i\vec f_2)\exp\{i(\beta z-\omega t)\}\right]$ and $\vec{\mathcal F}_2 = {\rm Im}\left[(\vec f_1+i\vec f_2)\exp\{i(\beta z-\omega t)\}\right]$. We can introduce a complex field $\vec{\mathcal F} = \vec{\mathcal F}_1+i\vec{\mathcal F}_2$, and define the phase $\varphi_\alpha(t)$ by choosing for some arbitrary point $(x_0,y_0,z_0)$ the value of the vector $\vec{\mathcal F}_{1,2}$ in a suitable direction, denoted by $F_{1,2}$. The phase is defined by $\exp\{i\varphi_\alpha(t)\} = (F_1(x_0,y_0,z_0,t)+iF_2(x_0,y_0,z_0,t))/|F(x_0,y_0,z_0,t)|$. In other words, $\varphi_\alpha$ is the phase of the complex vector field $\vec{\mathcal F}$ taken in some particular direction and at some particular point. 

The complex wave function $\psi_\alpha$ uses only part of the information of the classical electromagnetic field, since eq.~(\ref{eq:15A}) is an integral and the phase only involves a particular point and direction. In this sense our quantum system will be a subsystem of a classical probabilistic or deterministic system of microscopic classical fields. The practical realization of the field components $F_\tau$ in eq.~(\ref{eq:3}) is not unique. For example, one could use alternatively the mean of certain field components over the wave guide at a certain point $z$. An angle, and therefore a phase is defined by the ratio of such mean values.

For a bunch of isolated identical waveguides one has the simple evolution law
\begin{equation}
\psi(t+\varepsilon) = \exp(-i\omega\varepsilon)\psi(t)\,, \quad \bar H = \omega\,. \label{eq:16}
\end{equation}
An overall phase plays no role for a quantum system and we can subtract from $\bar H$ the constant $\omega$. This simple system has then a static evolution. We may also define the phase by evaluating $F$ for a time-dependent $z_0(t)$. For the particular choice $z_0(t) = \bar z_0 + \omega t/\beta$ the phase does not change, realizing directly $\bar H = 0$. This setting is most appropriate if instead of a static situation one considers an extended but finite light pulse.

The initial condition for the wave function is partly fixed by preparing input intensities for the different channels, determining $p_\alpha(t=0)$. In addition the initial wave function depends on relative phases between the different wave guides. These depend on the precise definition of the phase, e.g.\ the choice of $x_0$, $y_0$, the direction of $\vec{\mathcal F}$ or similar. Initial relative phases may be more difficult to control and we will come back to this issue later.

\section{Quantum gates}

A non-trivial quantum evolution can be realized if one introduces elements as beam splits, switches or phase shifts in certain regions for $z$. These elements have to be designed in order to be compatible with the chosen complex structure, such that the evolution continues to be described by the unitary step evolution operator $U(t)$ in eq.~(\ref{eq:12}). If we define the wave function by a time dependent $z_0(t)$ which increases monotonically with time, the change of the properties with $z$ will be mapped to a time dependence of $U(t)$. Non-trivial step evolution operators $U(t)$ can realize a quantum gate and sequences of quantum gates. One may adapt the time interval $\varepsilon$ such that it corresponds to the range $\Delta z$ for which a certain element (beam split etc.) is realized. The sequence of elements is then mapped to a sequence of unitary operators $U_i$ which in turn represent a sequence of quantum gates.

We may extend our setting by considering a wave function $\psi_\alpha(t,z)$ which depends on time and the position $z$ in the system of waveguides. One can then take the stationary limit where the form of the wave function remains the same for all times. Fixing a certain $t=t_0$ the wave function becomes a function of $z$. In dependence on $z$ the system realizes the same sequences of quantum gates $U_i$. This setting realizes the ``static memory materials'' discussed in ref.~\cite{Sexty2018}, with the interesting possible interplay of initial and final boundary conditions. In the present note we keep the focus on the setting close to a quantum computer. The initial conditions are prepared at the input position $z=0$ at a given time $t_{\rm in}=0$, and the result is read out at the output position of the system at $z_f$ at time $t_f$, which is determined by the time a light pulse needs to pass from $z=0$ to $z=z_f$. The sequence of quantum operators $U_i$ is the same for an evolution with $t$ or with $z$.

\subsection{Phase shift}

Realizing phase shifts in arbitrary channels $\alpha$ is a standard operation in modern photonics. It corresponds to a gate $U_P$ given by a diagonal matrix
\begin{equation}
U_P = {\rm diag}(e^{i\gamma_\alpha})\,, \label{eq:17}
\end{equation}
with $\gamma_\alpha$ the phase shift in the channel $\alpha$. For realizing a phase shift for a single qubit $j$ one needs to perform identical phase shifts in a number of channels. A phase shift of the $s_j=-1$ component of qubit $j$ changes the phases of all components of the wave function for which $s_j=-1$ by the same $\gamma_\alpha$, and leaves all components with $s_j=+1$ unchanged.

More formally, this can be seen in a direct product representation of the wave function
\begin{equation}
\psi = \psi_\alpha E_\alpha\,, \quad E_\alpha = E_1\otimes E_2\otimes\cdots\otimes E_{M_q}\,. \label{eq:18}
\end{equation}
Here $E_j$ can take the values $E_{j+} = \begin{pmatrix}1\\0\end{pmatrix}$ or $E_{j-} = \begin{pmatrix}0\\1\end{pmatrix}$. One takes $E_{j+}$ if for the labeling of basis vectors by a bit list as $(1,1,0,1,\dots,0)$ one has 1 at position $j$, and $E_{j-}$ if 0 stands at this position. (For the example with $M_q=3$ the basis vector $E_3$ corresponds to the bit list $(1,0,1)$ or $E_1=E_{1+}$, $E_2=E_{2-}$, $E_3=E_{3+}$.) A phase shift for the lower component of the qubit two corresponds in this representation to the direct product matrix
\begin{equation}
U = 1\otimes U_R \otimes 1 \otimes\cdots\otimes 1\,, \quad U_R = \begin{pmatrix}1&0\\0&e^{i\delta}\end{pmatrix}\,. \label{eq:19}
\end{equation}
This multiplies all components of the wave function for which a zero stands at the second position of the bit-list by a common phase. In our example with $M_q=3$ this implies $\gamma_3=\gamma_4=\gamma_7=\gamma_8=\delta$, $\gamma_1=\gamma_2=\gamma_5=\gamma_6=0$. For $\delta=\pi/4$ this simultaneous phase change realizes the standard rotation gate for the second qubit.

Rotation gates for single qubits are part of the basis gates for complex quantum operations. The rotation gate for an arbitrary qubit $j$ can be realized by a suitable combination of phase shifts for the channels $\alpha$. On the other hand a phase shift in a single channel $\alpha$ cannot be realized by a combination of phase shifts for individual qubits. We conclude that rotation gates for arbitrary qubits can be implemented, while in addition one has the possibility to perform operations acting simultaneously on several qubits by rather simple manipulations of individual channels $\alpha$. This is a direct consequence of a setting where the channels represent basis states of multi-qubit systems, rather than individual qubits. It underlines the difference of our setting to the realization of a qubit by a single channel or a single quantum photon.

\subsection{Switch}

Another simple operation switches two channels $\bar\alpha$ and $\bar\beta$. The field in channel $\bar\alpha$ is transported to channel $\bar\beta$, and vice versa. For a static realization one can simply cross the channels. For a programmable setting one may prefer a gate for which one can influence if the switch is performed or not. The unitary matrix $U_S$ describing the switch $\bar\alpha\leftrightarrow\bar\beta$ reads
\begin{equation}
U_{S;\alpha\beta}^{(\bar\alpha,\bar\beta)} = \delta_{\alpha\bar\alpha}\delta_{\beta\bar\beta} + \delta_{\alpha\bar\beta}\delta_{\beta\bar\alpha} + \delta_{\alpha\beta}(1-\delta_{\alpha\bar\alpha}-\delta_{\beta\bar\beta})\,. \label{eq:20}
\end{equation}
This is a generalization of $\tau_1$ to the case of more than two channels. For the example $M_q=3$ the switch $(\bar\alpha,\bar\beta)=(1,3)$ exchanges the components $(1,1,1)\leftrightarrow(1,0,1)$, leaving all other components invariant.

One can realize the CNOT gate by a suitable combination of switches. The CNOT gate acts on individual pairs of qubits. For example, the CNOT-gate for the qubits one and two is implemented in a direct product representation by
\begin{equation}
U_C^{(1,2)} = U_C\otimes 1\otimes 1\otimes\dots\,, \label{eq:21}
\end{equation}
with $U_C$ a $4\times4$ matrix acting on the first two factors,
\begin{equation}
U_C = \begin{pmatrix}1&0&0&0\\0&1&0&0\\0&0&0&1\\0&0&1&0\end{pmatrix}\,. \label{eq:22}
\end{equation}
For a two-qubit system it switches $(0,1)\leftrightarrow(0,0)$, corresponding to $U_S^{(3,4)}$. For a three qubit system $U_C^{(12)}$ does not change the state of the third qubit and therefore switches $(0,1,0)\leftrightarrow(0,0,0)$, $(0,1,1)\leftrightarrow(0,0,1)$. This is realized by the sequence of switches $U_S^{(5,7)}U_S^{(6,8)}$. Similarly, the CNOT-gate $U_C^{(2,3)}$ for qubits two and three is implemented by $(0,0,0)\leftrightarrow(0,0,1)$, $(1,0,0)\leftrightarrow(1,0,1)$ or $U_S^{(3,4)}U_S^{(7,8)}$. For four qubits the CNOT-gate for an arbitrary pair of qubits involves four switches, and for $M_q$-qubits one needs $2^{M_q-2}$ switches. CNOT-gates for arbitrary pairs of qubits can be implemented by suitable sequences of switches.

The pairs of channels involved in the switches needed for a CNOT-gate are disjunct in the sense that no pair involves a channel of another pair. These particular switches commute,
\begin{equation}
\left[U_S^{(\alpha,\beta)},U_S^{(\gamma,\delta)}\right] = 0 \quad\text{if}\quad \alpha\neq\gamma,\delta\,,\quad \beta\neq\gamma,\delta\,, \label{eq:23}
\end{equation}
such that the order of these switches does not matter. This property does not hold for general sequences of switches. The map of channels depends on the order of the switches, as for the example
\begin{align}
U_S^{(1,2)} U_S^{(1,3)} &: 3\to 2,\ 1\to 3,\ 2\to 1\,,\notag\\
U_S^{(1,3)} U_S^{(1,2)} &: 3\to 1,\ 1\to 2,\ 2\to 3\,. \label{eq:24}
\end{align}
General sequences of switches realize non-commuting unitary operations. Also the switches do not commute with the phase shifts.

It is well known that the CNOT-gate can transform direct product states for qubits into entangled states. This is also true for individual switches of channels. Every correlation $\langle s_a s_b s_c\dots\rangle$ is given by a linear combination of $\bar p_\alpha$ similar to eqs.~(\ref{eq:13}), (\ref{eq:14}). Switches can therefore be seen as maps in the space of correlations.

\subsection{Unitary beam split}

A beam split redistributes the intensities in two channels $\alpha,\beta$, $(I_\alpha,I_\beta)\to(I'_\alpha,I'_\beta)$. This distribution is such that if $I_\beta=0$ one has $I'_\alpha=I'_\beta=I_\alpha/2$. The beam $\alpha$ is split equally into the channels $\alpha$ and $\beta$. Similarly, we require for a unitary beam split that $I_\alpha=0$ results in $I'_\alpha=I'_\beta=I_\beta/2$. We suppose that the beam split is done in compatibility with the complex structure. It is then described by a unitary operation,
\begin{equation}
\begin{pmatrix}\psi'_\alpha\\\psi'_\beta\end{pmatrix} = U_B^{(\alpha,\beta)}\begin{pmatrix}\psi_\alpha\\\psi_\beta\end{pmatrix}\,, \label{eq:25}
\end{equation}
which leaves the other channels $(\gamma\neq\alpha,\beta)$ untouched. For the beam split our condition implies
\begin{equation}
U_B\begin{pmatrix}1\\0\end{pmatrix} = \frac{1}{\sqrt2}\begin{pmatrix}e^{i\gamma}\\e^{i\delta}\end{pmatrix}\,, \quad U_B\begin{pmatrix}0\\1\end{pmatrix} = \frac{1}{\sqrt2}\begin{pmatrix}e^{i\gamma'}\\e^{i\delta'}\end{pmatrix}\,, \label{eq:26}
\end{equation}
which results in
\begin{equation}
U_B = \frac{1}{\sqrt2}\begin{pmatrix}e^{i\gamma}&e^{i\gamma'}\\e^{i\delta}&e^{i\delta'}\end{pmatrix}\,. \label{eq:27}
\end{equation}
The condition for unitarity $U_B^\dagger U_B = 1$ imposes relations for the phases
\begin{equation}
e^{i(\delta-\delta')} = -e^{i(\gamma-\gamma')}\,. \label{eq:28}
\end{equation}
Up to phase shifts $U_B$ is the Hadamard gate $U_H$,
\begin{align}
U_B &= \begin{pmatrix}1&0\\0&e^{i(\delta-\gamma)}\end{pmatrix} U_H \begin{pmatrix}e^{i\gamma}&0\\0&e^{i\gamma'}\end{pmatrix}\,,\notag\\
U_H &= \frac{1}{\sqrt2}\begin{pmatrix}1&1\\1&-1\end{pmatrix}\,. \label{eq:29}
\end{align}
By a combination with suitable phase shifts every unitary beam split realizes the Hadamard gate between the involved channels $(\alpha,\beta)$. We will for simplicity assume that phases are chosen such that $U_B^{(\alpha,\beta)} = U_H$. For systems of more than one qubit the Hadamard gate $U_H^{(j)}$ acting on a single qubit $j$ can be constructed as a product of beam splits $U_B^{(\alpha,\beta)}$. The one-qubit Hadamard gate $U_H^{(j)}$ corresponds in the direct product representation to
\begin{equation}
U_H^{(j)} = 1\otimes\dots 1\otimes U_H\otimes 1\dots\otimes 1\,, \label{eq:30}
\end{equation}
with $U_H$ at the position $j$. For two qubits $U_H^{(1)}$ and $U_H^{(2)}$ correspond to the $4\times4$ matrices
\begin{align}
U_H^{(1)} &= \frac{1}{\sqrt2}\begin{pmatrix}1&0&1&0\\0&1&0&1\\1&0&-1&0\\0&1&0&-1\end{pmatrix}\,,\notag\\
U_H^{(2)} &= \frac{1}{\sqrt2}\begin{pmatrix}1&1&0&0\\1&-1&0&0\\0&0&1&1\\0&0&1&-1\end{pmatrix}\,. \label{eq:31}
\end{align}
On the other hand, one has for the beam splits of channels $(1,2)$ or $(3,4)$
\begin{equation}
U_B^{(1,2)} = \begin{pmatrix}U_H&0\\0&1\end{pmatrix}\,, \quad U_B^{(3,4)} = \begin{pmatrix}1&0\\0&U_H\end{pmatrix}\,. \label{eq:32}
\end{equation}
One obtains $U_H^{(2)}$ as the product of $U_B^{(1,2)}$ and $U_B^{(3,4)}$, $U_H^{(2)} = U_B^{(1,2)}U_B^{(3,4)} = U_B^{(3,4)}U_B^{(1,2)}$. Similarly, one finds $U_H^{(1)} = U_B^{(1,3)}U_B^{(2,4)}$. For systems with $M_q$ qubits the Hadamard gate for a single qubit $j$ is realized by a product of $2^{M_q-1}$ beam splits. These collective beam splits involve all channels, which are ordered in different pairs in order to realize the Hadamard gate for different qubits $j$. Hadamard-type gates typically change the direction of the spin of qubits. For a single qubit an eigenstate of $S_j^{(z)}$ is changed by $U_H$ to an eigenstate of $S_j^{(x)}$. Beam splits for a single pair of channels can be seen as maps in the space of correlations for quantum spins in different directions.

The practical construction of the unitary beam split with well controlled phases may constitute a challenge. While the evolution is naturally compatible with the complex structure for phase shifts and switches, this is not automatic for arbitrary forms of beam splits. If compatibility with the complex structure is not realized, the evolution is no longer described by a unitary step evolution operator. In this case the step evolution operator mixes the wave function with its complex conjugate. We do not pursue here the interesting question if operations beyond unitary operations, as for example the complex conjugation of the wave function, could open new computational possibilities.

\section{Quantum computer}

By suitable phase shifts and beam splits wave guides for classical electromagnetic fields can realize the rotation gate and the Hadamard gate for an arbitrary qubit $j$ within a system of $M_q$ qubits. With switches one can also implement the CNOT-gate for an arbitrary pair $(j,j')$ of two qubits. This realizes all basic quantum gates for a quantum computer. An arbitrary unitary transformation for the wave function of the $M_q$-qubit system can be performed by a suitable product of these basis gates.

In turn, this implies that we can also consider switches and beam splits between two channels $\alpha$ and $\beta$, together with phase shifts for individual channels $\alpha$, as a set of basis gates. They may be called ``correlation gates''. An arbitrary unitary transformation of the wave function for $M_q$ qubits can be performed by suitable products of correlation gates. Since these operations involve only two or one channel, they are easier to realize than the gates for individual qubits or pairs of qubits. Correlation gates may offer the advantage that a simple gate acts on many qubits at once.

The channels $\alpha$ encode directly the information about correlations for the spins of the qubits $s_i$. A sharp value $\bar p_\alpha=1$ corresponds to all intensity concentrated in a single channel $\alpha$. This fixes at once all expectation values of the qubits. For $M_q=3$ the state with $\bar p_3=1$ implies sharp values $\langle s_1\rangle=1$, $\langle s_2\rangle=-1$, $\langle s_3\rangle=1$, and therefore maximal values for all correlations $\langle s_1s_2\rangle=-1$, $\langle s_1s_3\rangle=1$, $\langle s_2s_3\rangle=-1$, $\langle s_1s_2s_3\rangle=-1$. More generally, each correlation is determined as a linear combination of channel probabilities (\ref{eq:14}). A switch between channels $\alpha=2$ and $\beta=3$ exchanges $\bar p_2\leftrightarrow\bar p_3$, leaving all other $\bar p_\gamma$ invariant. This changes at once $\langle s_2\rangle$, $\langle s_3\rangle$, $\langle s_1s_2\rangle$ and $\langle s_1s_3\rangle$ in dependence on the other probabilities $\bar p_\gamma$. On the level of qubits simple correlation gates correspond to conditional changes. So far quantum algorithms are mainly based on ``qubit gates'' involving one or two qubits, which can be realized in our setting as well. The correlation gates may offer additional possibilities.

For our correlation based quantum computer a transformation from a direct product state to an entangled state can be done easily.
For a direct product state the correlation between different qubits vanish, implying for $M_q=3$ the relations $\langle s_1 s_2\rangle=\langle s_1 s_3\rangle= \langle s_2 s_3\rangle=\langle s_1 s_2 s_3\rangle=0$.
This entails the conditions
\begin{align}
    &\bar p_5 = \frac{1}{4} - \bar p_4\,, \quad \bar p_6 = \frac{1}{4} - \bar p_3\,, \\
    &\bar p_7 = \frac{1}{4} - \bar p_2\,, \quad \bar p_8=\frac{1}{4} - \bar p_1\,, \\
    &\bar p_2 + \bar p_3  = \bar p_1 + \bar p_4\,,
    \label{eq:34A}
\end{align}
leaving only three independent $\bar p_\alpha$.
A simple switch $\bar p_1 \leftrightarrow \bar p_2$ violates these conditions and therefore transforms a direct product state into an entangled state.

For a quantum computer the number of qubits $M_q$ is a key quantity. In our setting the number of channels $2^{M_q}$ rises fast with $M_q$. It may be possible to realize many channels in a single wave guide by using different modes. Nevertheless, the technical possibilities will not allow very large $M_q$. It is an open question if a ``photonic quantum computer'' based on correlation gates can be useful for practical computations. In any case, it would be a practical demonstration that a quantum evolution can be realized by classical fields.

At this stage we have constructed a map from electromagnetic fields to a complex wave function, such that the Maxwell equations for optical circuits result in a discretized Schr\"odinger equation for sequences of quantum gates. This map may be useful, but it is conceptually not really surprising. After all, the Schr\"odinger equation can also be implemented by the manipulation of complex numbers in a classical von-Neumann computer, which is also based ultimately on electromagnetic fields. For a true quantum system one needs more. 
One has to implement observables which have a spectrum of possible measurement values which may be discrete, as for the quantum spin of a qubit in different directions. Not all of these observables have definite values for a given quantum state. Only a probability distribution for the measurement values is fixed by a quantum state. 
The quantum wave function encodes probabilistic information about the observables rather than being itself a directly observable object, as for our discussion of optical circuits. 
We discuss in the following probabilistic aspects of optical circuits which bring them even closer to real quantum systems.

\section{Probabilistic quantum computing}

It is nowadays possible to realize optical circuits with a rather large number of channels. Furthermore, one may use a large number of input states. This can be done in a sequential way by varying the input electromagnetic field. It may even be done in parallel using different frequencies in a range where the operation of the gates is frequency independent. It may no longer be feasible to control all details of the input states. One deals then rather with a probability distribution over possible input states. This will lead to some form of probabilistic quantum computing.

In particular, for a unitary evolution the phases of the wave function are important. A quantum computer based on classical fields needs a controlled evolution of the phases. If phase information is lost, the evolution is no longer unitary. On the other hand, it may sometimes be difficult to prepare input with well defined phases, or read out the phases in the output. One may consider a situation where only the input intensities $\bar p_\alpha(t=0)$ are known, while the phases of the initial wave function $\psi(t=0)$ are given by some probability distribution.  

For general probabilistic initial conditions our setting describes the unitary evolution of a quantum density matrix $\rho_{\alpha\beta}(t)$ according to the von Neumann equation $\rho(t+\varepsilon) = U(t)\rho(t)U^\dagger(t)$. If one has a probability distribution of initial wave functions, with $w_m$ the probability of a given initial wave function $\psi^{(m)}(t=0)$, the initial density matrix is given by
\begin{equation}
\rho_{\alpha\beta}(t=0) = \sum_m w_m \psi_\alpha^{(m)}(t=0)\psi_\beta^{(m)*}(t=0)\,. \label{eq:33}
\end{equation}
A known initial distribution of input intensities $\bar p_\alpha(t=0)$ fixes the diagonal elements $\rho_{\alpha\alpha}(t) = \bar p_\alpha(t)$.
The phase information is encoded in the off-diagonal elements. For our setting the extension to a probabilistic initial state is very straightforward. It is the same as for quantum mechanics. The wave function generalizes to the density matrix. The role of the unitary operations representing the gates remains unchanged. A density matrix contains additional probabilistic information beyond the wave function. This information concerns correlations between the basis states of the qubits. It may be used for computational purposes. It is not obvious if such a probabilistic setting can be achieved easily for a von-Neumann computer. 

Optical circuits with probabilistic initial conditions can realize type of a probabilistic quantum computer.
The microscopic evolution is given by the deterministic evolution of the electromagnetic fields according to Maxwell's equations. Only the input is probabilistic. In this sense one may view these systems as probabilistic cellular automata.

\section{Quantum observables}

In quantum mechanics a simple observable is the projector $P^{(\alpha)}$ on the channel $\alpha$, or associated basis state of the wave function. It is represented by a diagonal operator
\begin{equation}
\left(P^{(\alpha)}\right)_{\beta\gamma} = \delta_{\alpha\beta}\delta_{\alpha\gamma}\,. \label{eq:OP1}
\end{equation}
According to
\begin{equation}
\left(P^{(\alpha)}\right)^2 = P^{(\alpha)}\,, \label{eq:OP2}
\end{equation}
it has the eigenvalues one and zero. In quantum mechanics, these are the possible outcomes of measurements. For the optical circuit we can directly measure this quantum observable only for particular families of states. If the state has all intensity in the channel $\alpha$, $\bar I_\alpha = I_{\rm tot}$, one has $P^{(\alpha)}=1$, and if no intensity is found in the channel $\alpha$, $\bar I_\alpha=0$, one infers $P^{(\alpha)}=0$. For arbitrary states quantum mechanics predicts the expectation value of $P^{(\alpha)}$,
\begin{equation}
\langle P^{(\alpha)}\rangle = \psi_\alpha^*\psi_\alpha = I_\alpha/I_{\rm tot} = \bar p_\alpha\,, \label{eq:OP3}
\end{equation}
in the sense that measurements will find the values one or zero with a distribution yielding $\langle P^{(\alpha)}\rangle$. For the optical circuit eqs.~(\ref{eq:OP1}),~(\ref{eq:OP3}) can still be used. In contrast to quantum mechanics one can measure $I_\alpha/I_{\rm tot}$ and finds continuous values of this quantity.

We could push the analogy to quantum mechanics further by using a ``stochastic measurement apparatus''. The outcomes of this apparatus are only the two possibilities one and zero. The measurement apparatus ``reads'' $I_\alpha/I_{\rm tot}$ and uses a random generator to produce a distribution of outcomes according to the mean value $\langle P^{(\alpha)}\rangle$. This construction seems, however, rather ad hoc. In quantum mechanics this distribution arises naturally. The important difference is not modified substantially if one employs probabilistic initial conditions.
For the purpose of constructing a quantum computer based on optical circuits the ``discretization of the output'' may not be of major importance.

The situation for the quantum spin $S_j^{(z)}$ of the qubit $j$ is similar. For optical circuits one can extract the expectation value $\langle S_j^{(z)}\rangle$ by the quantum rule from the wave function. For optical circuits the sharp values $\pm1$ of quantum mechanics will be found for particular linear combinations of $I_\alpha$ according to eq.~(\ref{eq:13}). For arbitrary states the discretization to the possible measurement values $\pm1$ we may proceed again by a stochastic measurement apparatus.

What about measurements of the quantum spin in a direction different from $S_z$? We may exemplify this by the measurement of $S_x$ in case of a single qubit. A measurement apparatus for $S_x$ can be built by performing two steps: 1) Execute a Hadamard gate by a beam split, and 2) measure $S_z$ by comparing intensities in the two channels $\alpha=1$ and $\alpha=2$. The second step may be done by a stochastic measurement apparatus. As a consequence, this apparatus measures $\langle S_z\rangle$ in the state $U_H\psi$,
\begin{equation}
\langle S_x\rangle = \psi^\dagger U_H^\dagger S_z U_H \psi\,. \label{eq:OP4}
\end{equation}
Indeed, if $\psi_+$ is an eigenstate of $S_x$ with eigenvalue one, the Hadamard transformed state $U_H\psi_+$ is an eigenstate of $S_z$ with eigenvalue one,
\begin{equation}
\psi_+ = \frac{1}{\sqrt2}\begin{pmatrix}1\\1\end{pmatrix}\,, \quad U_H\psi_+ = \begin{pmatrix}1\\0\end{pmatrix}\,. \label{eq:OP5}
\end{equation}
This holds similarly for the eigenstate with eigenvalue $-1$,
\begin{equation}
\psi_- = \frac{1}{\sqrt2}\begin{pmatrix}1\\-1\end{pmatrix}\,, \quad U_H\psi_- = \begin{pmatrix}0\\1\end{pmatrix}\,. \label{eq:OP6}
\end{equation}
If the stochastic measurement apparatus for $S_z$ produces after the beam split the value $+1$ with a probability $p_+$, and $-1$ with a probability $p_- = 1-p_+$, one infers $\psi^\dagger U_H^\dagger S_z U_H\psi = p_+-p_-$. For a general state
\begin{equation}
\psi = \alpha_+\psi_+ + \alpha_-\psi_-\,, \quad U_H\psi = \begin{pmatrix}\alpha_+\\\alpha_-\end{pmatrix}\,, \label{eq:OP7}
\end{equation}
this implies
\begin{equation}
p_+ = |\alpha_+|^2\,, \quad p_- = |\alpha_-|^2\,. \label{eq:OP8}
\end{equation}
Therefore the measurement apparatus for $S_x$ will yield the value $+1$ with probability $|\alpha_+|^2$, and $-1$ with probability $|\alpha_-|^2$. This is precisely the quantum rule for the expectation value $\langle S_x\rangle$,
\begin{equation}
\langle S_x\rangle = |\alpha_+|^2 - |\alpha_-|^2\,. \label{eq:OP9}
\end{equation}
If we associate to $S_z$ the operator $\hat S_z = \tau_3$, the operator for $S_x$ is given by $\tau_1$,
\begin{equation}
\hat S_z = \tau_3\,, \quad \hat S_x = U_H^\dagger \hat S_z U_H = \tau_1\,. \label{eq:OP10}
\end{equation}
The operators $\hat S_x$ and $\hat S_z$ do not commute. This entails the quantum-mechanical uncertainty relations. A measurement of $S_y$ is similar, but accompanying the beam split by suitable phase shifts. The generalization to several qubits is straightforward.

This raises questions about ``no-go theorems'' for the embedding of quantum mechanics in classical statistics, based on Bell's inequalities \cite{Bell1964,Clauser1969}. We can treat the spins $S_j^{(z)}$ as classical Ising spins $s_j$.
If we associate the relative intensities $I_\alpha/I_{\rm tot}$ to probabilities by the use of a stochastic measurement apparatus, the subsystem defined by the probabilistic information in the wave function $\psi$,  contains enough information for the computation of a probability distribution for configurations of the Ising spins, given by $\bar p_\alpha$. This allows the computation of classical correlation functions for the Ising spins. They have to obey Bell's inequalities. At this level no problem arises since the qubit-spins $S_j^{(z)}$ are all in the same $z$-direction and therefore commute.

What about other observables as the quantum spins $S_j^{(x)}$, $S_j^{(y)}$ in other directions? The associated quantum operators do not commute with the ones for $S_j^{(z)}$. Then certain quantum correlations violate Bell's inequalities. One concludes that the quantum correlation $\langle S_j^{(x)}S_j^{(z)}\rangle_q$ cannot be a classical correlation function. Indeed, the expectation values of $S_j^{(x)}$ and quantum correlations of $S_j^{(x)}$ with $S_j^{(z)}$ cannot be extracted from the probabilities $\bar p_\alpha$. They can be computed from the wave function $\psi_\alpha$, but the phases play a role. More generally, the determination of $\langle S_j^{(x)}\rangle(t)$ involves the density matrix $\rho_{\alpha\beta}(t)$. Beyond the diagonal elements the off-diagonal elements of $\rho$ matter. On the level of the subsystem there exists no probability distribution which determines simultaneous probabilities for $S_j^{(z)}$ and $S_j^{(x)}$. Thus no classical correlation function exists for this pair of observables, and Bell's inequalities for classical correlation functions do not apply.

On the other hand, one can compute the wave function $\psi(t)$ from the full information in the electromagnetic field. Thus the full electromagnetic field determines also the quantum correlations as $\langle S_j^{(x)}S_j^{(z)}\rangle_q$. For a given probabilistic distribution of electromagnetic fields one may further define some classical correlation function $\langle S_j^{(x)}S_j^{(z)}\rangle_{\rm cl}$, which has to obey Bell's inequalities. Such classical correlation functions differ from the quantum correlation. They depend on details of the definition and cannot be computed from the probabilistic information in the subsystem encoded in $\rho_{\alpha\beta}(t)$ \cite{ProbabilisticWorld}.

From the point of view of the subsystem the observables $S_j^{(x)}$, $S_j^{(y)}$ are statistical observables. They describe properties of the probabilistic information in the subsystem. If we define subsystem states by configurations of Ising spins $s_j$, the observables $S_j^{(x,y)}$ do not take fixed values in these states. Thus classical correlations for these observables do not exist within the subsystem. On the other hand, the wave function $\psi_\alpha(t)$ or the density matrix $\rho_{\alpha\beta}(t)$ contains sufficient information for the computation of expectation values of observables beyond the Ising spins $s_j$ and correlations $s_is_j$ etc.\ at a given time $t$. In our quantum formalism the Ising spins and their correlations are represented by diagonal operators. Additional observables correspond to non-diagonal operators, which no longer commute with the ones for the Ising spins.

What would a single quantum photon do if one uses this as input for the system of wave guides? Without a detailed investigation one can guess that the outcome to find it at $t_f$ in a given channel $\alpha$ is given by $p_\alpha(t_f)$. In this case the single photon replaces the stochastic device, with yes/no questions asking if the photon detector for channel $\alpha$ fires or not. 
A single quantum photon realizes a true quantum system, to be compared with our ad hoc construction of a stochastic measurement apparatus.
For the purpose of this note we do not need to explore the single photon case with its technical challenges. Classical electromagnetic fields are sufficient for the quantum evolution of the wave functions or density matrix.

\section{Discussion}

Classical electromagnetic fields can realize a quantum evolution. We have constructed a map from electromagnetic fields in wave guides to the complex wave functions for $M_q$ qubits. Phase shifts, switches and beam splits can realize quantum gates for these qubits, implementing arbitrary unitary evolution steps for the wave function. The corresponding correlation based photonic quantum computer is not based on a few isolated photons representing qubits. The channels of wave guides rather represent the $2^{M_q}$ basis states for a system of $M_q$ qubits in terms of classical fields. The increase of the number of needed channels $\sim 2^{M_q}$ sets technical limits for the size of this type of photonic quantum computer.

Simple correlation gates acting only on one or two channels affect many qubits at once, offering perhaps new possibilities for quantum algorithms. In addition, collective phase shifts, switches or beam splits acting simultaneously on all or a large number of channels realize operations on single qubits or pairs of qubits. One observes a duality: operations on one or two qubits affect simultaneously many basis states, while operations on one or two channels or basis states affect simultaneously many qubits. Usual quantum computers do not have simple possibilities to realize transformations involving only one or two basis states. The simple operations on basis states of a correlation based quantum computer may offer additional computational possibilities. This could partly compensate for the limited number of qubits.

We emphasize that all features of the quantum evolution are realized in our setting. Entanglement between different qubits is no problem. If all intensity is concentrated in a single channel $\alpha$, this represents a basis state which involves all qubits. It is a very highly entangled state. Interference and the superposition principle are inherited from the linearity of Maxwell's equations. Probabilistic initial conditions lead to a quantum density matrix in a rather straightforward way. 

As far as the realization of a true quantum system is concerned, possible issues are not related to the quantum evolution. 
They rather concern the probabilistic nature of the outcome of measurements. 
One may implement this by a ``stochastic measurement apparatus''. 
This is rather a demonstration of existence and may seem not very satisfactory from the point of view of an explanation of fundamental quantum mechanics. 
For known ``true'' quantum systems the discreteness of certain observables is implemented by suitable quantum axioms and the wave function is not directly observable.
For the optical circuits with classical fields the discrete observables require particular properties of the measurement apparatus, while the wave function can, in principle, be measured.

In the other direction our proposal develops the language of a quantum description for the time evolution in a large family of optical circuits. This may help a better understanding of certain aspects of existing circuits. With the general ansatz of refs.~\cite{Wetterich2018b,ProbabilisticWorld} this approach may extend to a large class of systems with dynamical electromagnetic fields, including a stochastic environment.

At this stage one may compare the correlation based photonic quantum computer with a von-Neumann computer. The components of the wave function are complex numbers, and the quantum gates are simple linear operations on complex numbers. This can be done, of course, by a von-Neumann computer as well. Advantages of a photonic version could be speed and reduced energy consumption. Furthermore, optical circuits seem well adapted to deal with probabilistic initial conditions. They can implement rather naturally the quantum evolution of a density matrix.

The present work may open new perspectives on a wider class of probabilistic or correlated computing. One may drop the compatibility with the complex structure. The step evolution operator for the real wave function is then orthogonal, generalizing the unitary evolution. For conserved total intensity the hermitian Hamilton operator is replaced by an antisymmetric operator $W$ \cite{ProbabilisticWorld}. We may call this generalization of quantum mechanics ``$W$-dynamics'', with $Q$-dynamics arising in the presence of a complex structure which is compatible with the evolution.

The map of classical fields to probability amplitudes opens a large variety of simulations of probability distributions for which no information is lost in the course of their time evolution. We have discussed already the case for which the evolution is deterministic, while initial conditions or input are probabilistic. We can also consider the possibility that the gates, for example the beam splits, act in a stochastic way. The ``phases'' of gates can be random. As long as the total intensity is preserved and the transformation remains linear, this is a random rotation of a real wave function. In contrast to standard versions of a probabilistic evolution of probabilities in Markov chains, the orthogonality and linearity of the evolution steps for wave functions preserves the information. The system needs not approach an equilibrium state for large time.

Crucial features of quantum dynamics, as the superposition principle and interference for wave functions, remain valid for stochastic $W$-dynamics. For all forms of probabilistic or stochastic computing one can further use the information stored in the correlations for the field values. Finally, the electromagnetic fields in the input and output channels contain more information than the components of the wave functions $q_\tau(t)$ discussed here.

There are different models which can be simulated by the classical electromagnetic fields of optical circuits. For example, one could take a system of $M_q$ fermions, with fermions of different types $j$ either present ($n_j=1$) or not ($n_j=0$). The dynamics may be characterized by microscopic (or hidden) parameters, corresponding to the microscopic electromagnetic field. They determine at every time $t$ the probability distribution $\{\bar p_\alpha(t)\}$, or more generally the density matrix $\rho_{\alpha\beta}(t)$, for the configurations of present or absent fermions (corresponding to ones or zeros in the bit list for $\alpha$). Suppose that the evolution of the expectation values follows the quantum law according to the correlation based photonic computer. 
For a stochastic measurement apparatus the possible measurement value for a fermion present or not is $n_j$, which can take the values one or zero. 
Its expectation value is then given by the quantum law $\langle n_j\rangle = \psi^\dagger \hat n_j \psi$, with operator $\hat n_j$ in the direct product basis $\hat n_j = 1\otimes 1\cdots\otimes N_j\otimes\cdots\otimes 1$, $N_j = (\tau_3+1)/2$. 
The fermionic system evolves according to quantum mechanics. We do not want to propose this as a model for real particles. 
A more realistic route towards real quantum particles emerging from classical probabilistic systems is given by probabilistic cellular automata \cite{Wetterich2021,Wetterich2022,Kreuzkamp2024} or subsystems of classical field theories with probabilistic initial conditions \cite{CWL,CWDS}.
Nevertheless, optical circuits may be useful to simulate quantum systems for fermions if the number of degrees of freedom is not too large.

\bibliography{refs}

\end{document}